\documentclass[journal=jpcbfk, manuscript=article, layout=twocolumn, doi=true]{achemso}

\usepackage[utf8]{inputenc}
\usepackage{amsmath}
\usepackage{amssymb}
\usepackage{graphicx}
\graphicspath{{figures/}}
\usepackage{physics}
\usepackage{booktabs}
\usepackage{xcolor}
\usepackage{subcaption}
\usepackage{algorithm}
\usepackage{algpseudocode}
\usepackage{float}

\title{Non-Equilibrium Dynamics of the Time-Dependent Excitonic Coupling in Fluorescent Protein Dimers}

\author{Robson Christie}
\affiliation{School of Computing, Mathematics and Physics,
University of Portsmouth, Portsmouth PO1 3FX, U.K.}

\author{Cerys Murray}
\affiliation{School of Mathematics and Physics, University of Surrey, Guildford GU2 7XH, U.K.}

\author{Youngchan Kim}
\affiliation{Leverhulme Quantum Biology Doctoral Training Centre, University of Surrey, Guildford GU2 7XH, U.K.; School of Biosciences, University of Surrey, Guildford GU2 7XH, U.K.; Advanced Technology Institute, University of Surrey, Guildford GU2 7XH, U.K.}

\author{Jaewoo Joo}
\email{* jaewoo.joo@port.ac.uk}
\affiliation{School of Computing, Mathematics and Physics,
University of Portsmouth, Portsmouth PO1 3FX, U.K.}

\begin{document}

\maketitle

\begin{tocentry}
\includegraphics[width=1.0\textwidth]{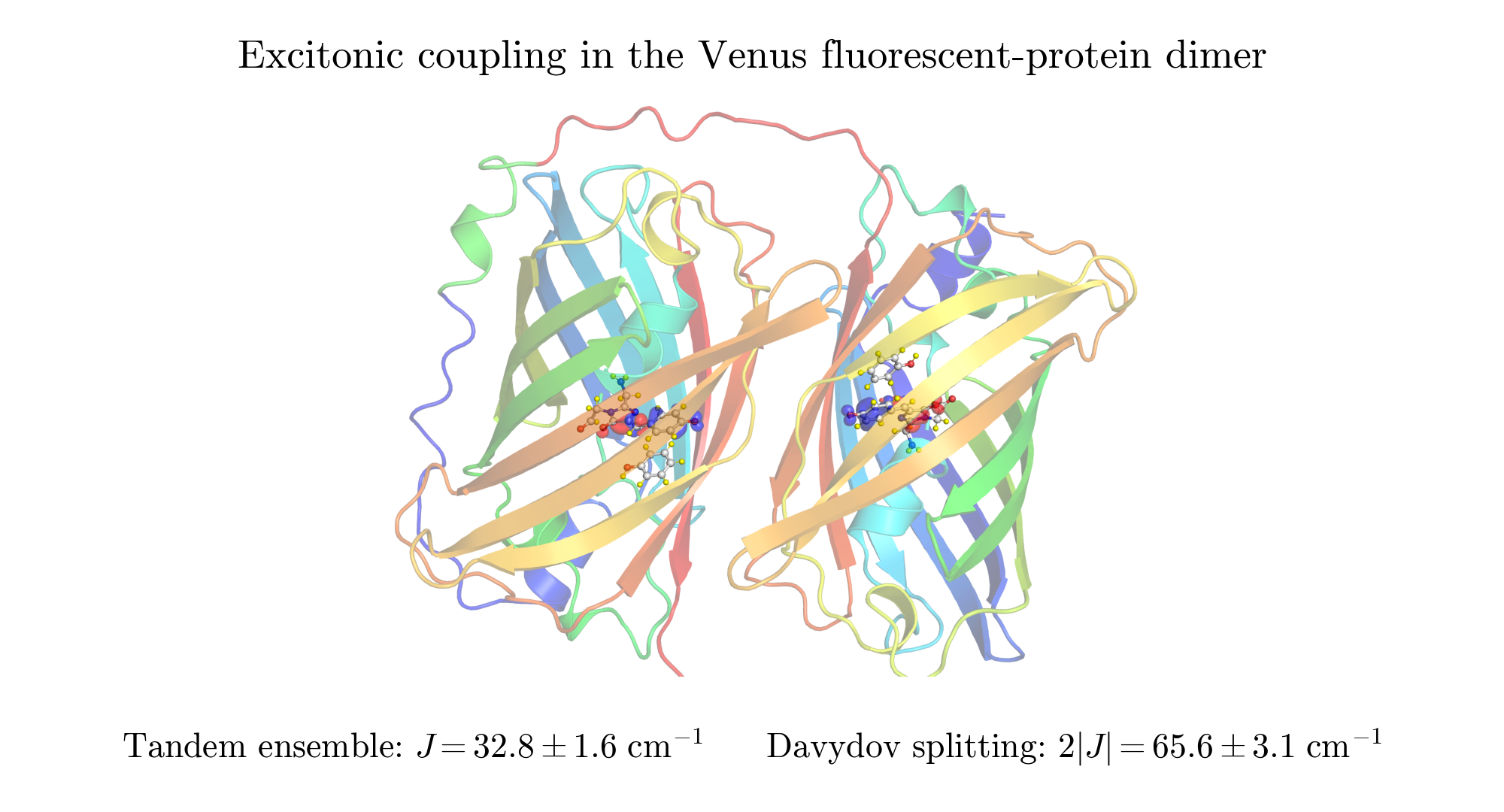}
\end{tocentry}

\begin{abstract}
We quantify the excitonic coupling in the homodimer of the dimeric Venus fluorescent protein using a quantum-classical hybrid workflow. Because the anionic chromophore carries appreciable double-excitation character, we obtain its site energy from domain-based local-pair-natural-orbital similarity-transformed equation-of-motion coupled-cluster theory (DLPNO-STEOM-CCSD), which places the in-protein bright $\pi\pi^{*}$ state at $523.90~\mathrm{nm}$, in good agreement with the experimental absorption maximum near $515~\mathrm{nm}$ and validated against an explicit triples-corrected EOM-CCSD(fT) benchmark; single-excitation time-dependent density functional theory (TDDFT) blueshifts this state. Applying a transition-density coupling (TDC) formalism to the STEOM transition density and simulating the VenusA206 tandem-dimer construct (two $\beta$-barrels covalently joined by an inter-domain linker), the covalent tether holds the A206 interface docked throughout $1~\mathrm{ns}$ of room-temperature dynamics. We obtain a thermally averaged coupling of $J = 32.8 \pm 1.6~\mathrm{cm^{-1}}$ across the $1000$-frame trajectory (Davydov splitting $2|J| = 65.6 \pm 3.1~\mathrm{cm^{-1}}$), representing a $19\%$ near-field enhancement over the ensemble point-dipole estimate ($J_{\mathrm{PDA}} = 27.6 \pm 1.3~\mathrm{cm^{-1}}$) at the $24.69 \pm 0.32~\text{\AA}$ chromophore centroid separation. Furthermore, we argue that a separation of timescales resolves the apparent theoretical tension between intermediate experimental excitonic couplings and the highly decoherent biological environment. While it has been hypothesised that the fluorescent protein $\beta$-barrel scaffold sustains coupling by shielding the chromophore from thermal fluctuations, we emphasise that the separation of timescales applies irrespective of the exact degree of environmental decoherence and dissipation. Collective photoexcitation imprints the Davydov splitting under optical-limit dielectric screening upon absorption, preceding bulk solvent relaxation and sub-picosecond environmental dephasing. To characterise the subsequent post-absorption evolution, we employ stochastic wavefunction simulations to model the transition from a delocalised exciton superposition to incoherent hopping between localised chromophore states.
\end{abstract}

\section{Introduction}
Excitation energy transfer in biological systems is conventionally understood through the competition between intermolecular excitonic coupling ($J$) and environmental vibrational dephasing \cite{forster1965delocalized, may2004charge}. When environmental fluctuations dominate, phase relationships randomise rapidly, restricting the system to the very weak-coupling regime characterised by incoherent Förster hopping \cite{forster1948intermolecular}. Conversely, the intermediate weak regime permits the formation of coherently delocalised excitonic eigenstates, manifesting spectroscopically as Davydov splitting \cite{davydov1971}, while the strong coupling regime gives rise to more pronounced modifications of the absorption spectrum \cite{clegg2006history}. Because physiological environments are highly dynamic and structurally disordered, biophysics has historically assumed that ambient thermal noise precludes robust excitonic delocalisation, confining such systems to the very weak coupling regime \cite{clegg2006history, nelson2018role}.

Homodimers of the yellow fluorescent protein variant dimeric Venus challenge this assumption. Operating at room temperature, these dimers exhibit distinct Davydov splitting in circular dichroism (CD) spectra, and photon antibunching statistics confirm that the interacting chromophores act as a collective quantum entity upon photon absorption \cite{kim2019venusa206, freed2025effect}. These signatures indicate an interaction strength firmly within the intermediate excitonic regime. Couplings of this magnitude, and substantially larger, are documented in photosynthetic pigment--protein complexes~\cite{van2000photosynthetic, adolphs2006proteins, krueger1998} (for example, ${\sim}100~\mathrm{cm^{-1}}$ between neighbouring bacteriochlorophylls in the FMO complex and ${\sim}300~\mathrm{cm^{-1}}$ in the LH2 B850 ring). In the FMO complex, however, this exciton and Davydov structure is resolved in linear absorption and circular-dichroism spectra recorded at $6~\mathrm{K}$, where homogeneous broadening is suppressed~\cite{adolphs2006proteins}. The distinguishing feature of the Venus dimer is therefore not the magnitude of $J$ but the persistence and spectral resolvability of its coherent excitonic signatures at \emph{room} temperature, in an environment otherwise expected to induce strong decoherence~\cite{ishizaki2009unified}.

To explain this phenomenon, it has been hypothesised that the fluorescent protein $\beta$-barrel scaffold plays a central role in sustaining inter-chromophore excitonic coupling by providing a structured dielectric shield that attenuates thermal fluctuations from the surrounding environment, thereby potentially extending excitonic decoherence times \cite{kim2019venusa206}. While this protective shielding offers a compelling mechanism, we propose that the {persistence of the} excitonic coupling {signatures} can be reconciled by separating the physical timescales of exciton formation and subsequent dephasing, irrespective of the exact degree of environmental noise suppression. Photoexcitation acts as a collective, near-instantaneous vertical transition directly into a delocalised bright eigenstate, initially protected by the scaffold's optical-limit dielectric screening. The Davydov splitting is an inherent property of the energy spectrum imprinted at this precise moment of absorption. As demonstrated by the stochastic trajectories in this work (see {\bf Figure \ref{fig:ExcitonDynamics}}), the system subsequently undergoes strong environmental dephasing and rapid relaxation into a localised state. However, because this decoherence and the associated dielectric screening occur on sub-picosecond to picosecond timescales, they are significantly slower than the absorption event. Consequently, strong excitonic coupling and its spectroscopic signatures are observed despite the presence of strong, subsequent decoherence.

Accurately predicting the magnitude of the coupling that drives this initial delocalisation requires examining the limitations of standard point-dipole interaction models. At the inter-chromophore separation in dimeric Venus ($\sim 24.7$~\text{\AA}), the point-dipole approximation (PDA) provides a reasonable first-order estimate of the Coulombic interaction, but higher-order near-field multipolar contributions provide a modest correction to the overall coupling strength $J$ \cite{scholes2003, kitohnishioka2017forster}. That the PDA begins to break down once the inter-chromophore separation becomes comparable to the spatial extent of the transition density is well established~\cite{sobakinskaya2018theory, scholes2003}; our aim is to quantify the magnitude of this effect for the Venus dimer and, through an explicit Cartesian multipole decomposition, to evaluate how higher multipolar terms contribute. Capturing these effects rigorously requires integrating the full three-dimensional (3D) spatial distribution of the electron transition density, which maps the shape of the transition's probability amplitude onto the 3D molecular structure.

Furthermore, because exciton formation precedes the macroscopic nuclear reorganisation of the solvent (Debye relaxation $\tau_D \approx 8.3$\,ps), the initial Coulombic coupling is not damped by the static permittivity of water ($\varepsilon_{\mathrm{eq}} \approx 78$). Instead, it is screened only by the fast electronic polarisation of the environment, corresponding to the optical dielectric limit ($\varepsilon_{\mathrm{opt}} \approx 1.78$) \cite{mennucci2012, scholes2003}.

Obtaining a quantitatively reliable site energy and transition density is itself non-trivial for this system: the deprotonated (anionic) chromophore of yellow fluorescent proteins carries appreciable double-excitation character, precisely the regime in which TDDFT, a single-excitation theory, is known to fail, systematically blueshifting the bright state relative to correlated wavefunction methods and experiment. We therefore compute the site energy with DLPNO-STEOM-CCSD, a correlated equation-of-motion coupled-cluster method whose treatment of double (and, in part, triple) excitations we outline in the Theoretical Framework [Eq.~(\ref{eq:ci_expansion})] and detail in the Methods, and validate against an explicit triples-corrected EOM-CCSD(fT) benchmark, evaluating the coupling from the resulting STEOM transition density.

In this work, we present a quantum-classical hybrid analysis of non-equilibrium dynamics of the time-dependent excitonic coupling between dimeric Venus molecules. By combining a transition-density coupling (TDC) formalism with a non-equilibrium dielectric framework, we quantify the near-field multipolar contributions to the excitonic coupling. We demonstrate that the interplay of optical-limit screening and transition-density geometry produces an intermediate coupling ($J = 32.8 \pm 1.6~\mathrm{cm^{-1}}$, Davydov splitting $2|J| = 65.6 \pm 3.1~\mathrm{cm^{-1}}$), and show how non-equilibrium dynamics capture the subsequent decoherence and state localisation.
\section{Theoretical Framework}

\begin{figure*}[t]
  \centering
  \begin{subfigure}[b]{0.32\textwidth}
    \centering
    \includegraphics[width=\linewidth]{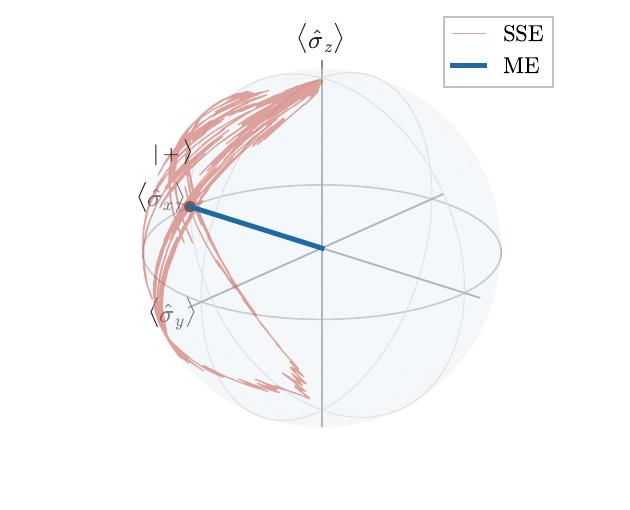}
    \caption{}
    \label{fig:bloch}
  \end{subfigure}\hfill
  \begin{subfigure}[b]{0.32\textwidth}
    \centering
    \includegraphics[width=\linewidth]{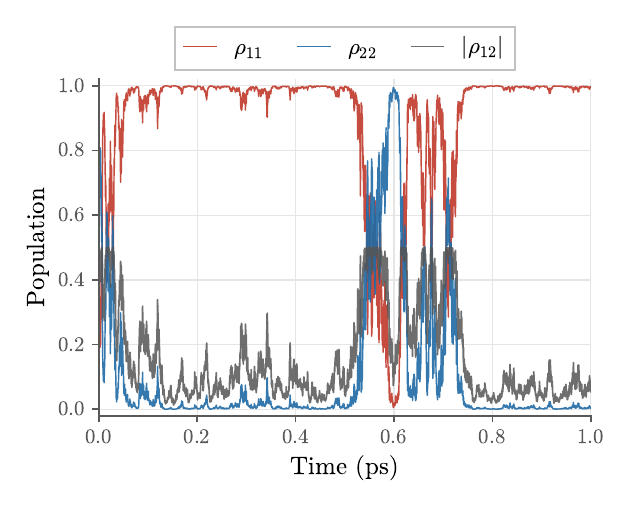}
    \caption{}
    \label{fig:sse_site}
  \end{subfigure}\hfill
  \begin{subfigure}[b]{0.32\textwidth}
    \centering
    \includegraphics[width=\linewidth]{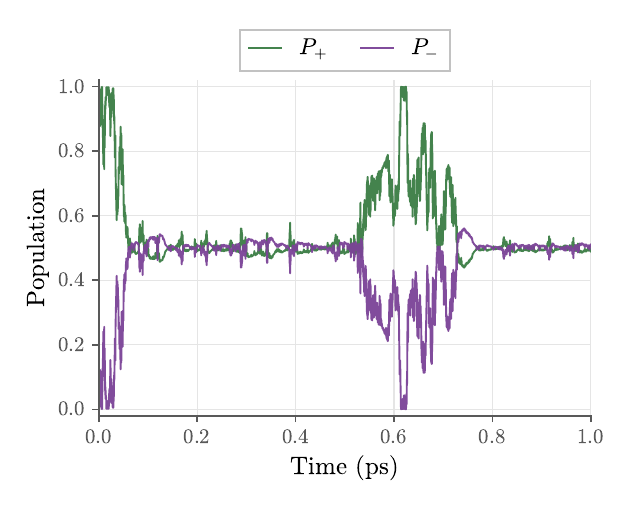}
    \caption{}
    \label{fig:sse_adiabatic}
  \end{subfigure}
  
  \vspace{0.3cm}
  
  \begin{subfigure}[b]{0.32\textwidth}
    \centering
    \includegraphics[width=\linewidth]{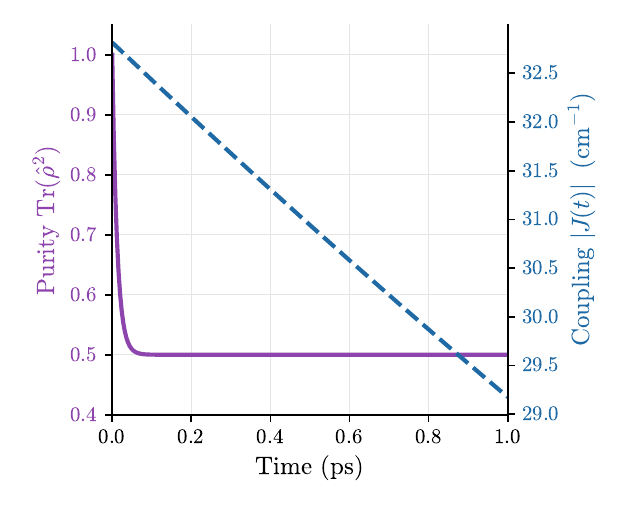}
    \caption{}
    \label{fig:purity_coupling}
  \end{subfigure}\hfill
  \begin{subfigure}[b]{0.32\textwidth}
    \centering
    \includegraphics[width=\linewidth]{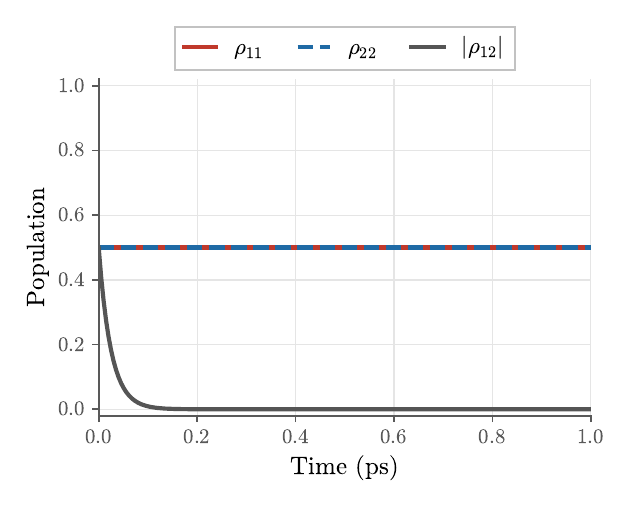}
    \caption{}
    \label{fig:me_site}
  \end{subfigure}\hfill
  \begin{subfigure}[b]{0.32\textwidth}
    \centering
    \includegraphics[width=\linewidth]{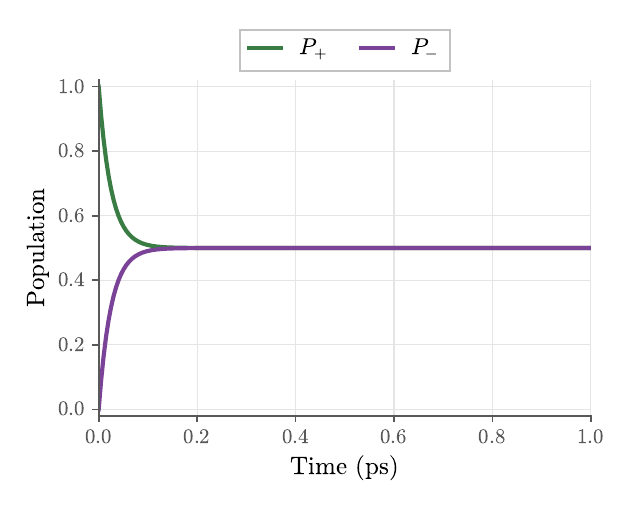}
    \caption{}
    \label{fig:me_adiabatic}
  \end{subfigure}

  \caption{ 
  Simulation of excitonic dynamics in the Venus dimer with the local site states $\ket{1}\equiv\ket{e_L g_R}$ and $\ket{2}\equiv\ket{g_L e_R}$: (a) Unified Bloch sphere showing the stochastic pure-state trajectory (red) and ensemble-averaged mixed-state path (blue); (b) Site-basis populations and coherence for a single SSE trajectory; (c) Adiabatic populations of bright and dark states $\ket{\pm}$ for a single SSE trajectory. (d) Evolution of the purity $\mathrm{Tr}(\hat{\rho}^2)$ and the magnitude of the time-dependent excitonic coupling $|J(t)|$; (e) Ensemble-averaged Master Equation (ME) site-basis populations and coherence; (f) Ensemble-averaged ME adiabatic populations. Time progression: in (a) both paths begin at the bright state $\ket{+}$ on the equator, with the individual trajectory (red) executing surface-preserving stochastic hops and the ensemble average (blue) contracting radially inward toward the maximally mixed centre; in (b-f) time increases from left to right. The rapid sub-$100$\,fs decay of the site-basis coherence $|\rho_{12}|$ is shown explicitly in panels (b) and (e).}
  \label{fig:ExcitonDynamics}
\end{figure*}

To describe the excitonic interactions within the dimer, we restrict our model to the single-exciton manifold, defining the local site basis as $\ket{1}\equiv\ket{e_L \, g_R}$ and $\ket{2}\equiv\ket{g_L \, e_R}$. This notation explicitly enforces the conservation of energy in a single-photon absorption subspace, ensuring that only a single excitation is shared between the two chromophores. The $\beta$-barrel of Venus has a diameter of $\sim 22$~\AA\ and a closest inter-chromophore approach of 15~\AA\ in the crystal structure of the dimeric Venus fluorescent protein homodimer~\cite{kim2019venusa206}, effectively precluding significant short-range Dexter exchange interactions~\cite{skourtis2016dexter}. Consequently, the electronic coupling $J$ is accurately described as the Coulomb interaction between the transition densities of the individual monomers \cite{hsu2009}. The electron transition density, $\rho^{g\to e}_m(\mathbf r)$, is obtained by integrating the product of the ground and excited state wavefunctions over the coordinates of $N-1$ electrons in a monomer:
\begin{multline} \label{eq:transition_density}
    \rho^{g\to e}_m(\mathbf{r}) =  
    N \int \dots \int d\mathbf{r}_2 \dots d\mathbf{r}_N \\ \Psi_e^*(\mathbf{r}, \mathbf{r}_2, \dots, \mathbf{r}_N) \, \Psi_g(\mathbf{r}, \mathbf{r}_2, \dots, \mathbf{r}_N)
\end{multline}
where $\Psi_g$ and $\Psi_e$ are the many-body electronic wavefunctions for the ground and excited states, respectively ($m=L,\,R$). 
The excited-state wavefunction $\Psi_e$ is itself a superposition of configurations obtained by promoting electrons out of the orbitals occupied in the ground-state reference $\Phi_0$ into unoccupied (virtual) orbitals,
\begin{multline} \label{eq:ci_expansion}
    \Psi_e = c_0\,\Phi_0 + \sum_{ia} c_i^{a}\,\Phi_i^{a}
    + \sum_{i<j,\,a<b} c_{ij}^{ab}\,\Phi_{ij}^{ab} \\
    {}+ \sum_{i<j<k,\,a<b<c} c_{ijk}^{abc}\,\Phi_{ijk}^{abc} + \cdots,
\end{multline}
where $i,j,k$ label occupied and $a,b,c$ virtual orbitals: $\Phi_i^{a}$ promotes one electron (a \emph{singly} excited, one-particle/one-hole configuration), $\Phi_{ij}^{ab}$ promotes two at once (a \emph{doubly} excited, ``2p2h'' configuration) and $\Phi_{ijk}^{abc}$ three (\emph{triply} excited, ``3p3h''). Single-excitation theories (TDDFT and configuration-interaction singles) retain only the $c_i^{a}$ amplitudes; when the true state carries appreciable weight in the doubly excited $c_{ij}^{ab}$ block, as the anionic Venus chromophore does, they misplace it and, here, blueshift the bright $\pi\pi^{*}$ state.

We therefore obtain $\Psi_e$ from DLPNO-STEOM-CCSD, a correlated coupled-cluster method that recovers the full double-excitation character, together with a partial subset of the triple excitations, that a single-excitation theory omits (the STEOM construction is summarised in the Methods). It thereby approaches the accuracy of an explicit full-triples treatment at a small fraction of the cost, which is decisive here: an explicit-triples calculation of the in-protein chromophore, embedded in the protein and solvent point-charge field, is computationally out of reach. We instead validate the method on the smaller, tractable gas-phase chromophore, where an explicit non-iterative triples correction [EOM-CCSD(fT)] can be afforded: STEOM-CCSD reproduces the triples-corrected excitation energy and removes the overestimate of the bare doubles-level EOM-CCSD (Table~\ref{tab:ladder}), confirming that it captures the higher-order excitation physics governing this transition.

Crucially, while the total electron density is a positive-definite observable representing the probability of finding an electron, the electron transition density implies an off-diagonal matrix element. Since $\int \rho^{g\to e}(\mathbf{r}) d\mathbf{r} = 0$ due to the orthogonality of the ground and excited states, its spatial distribution, characterised by positive and negative phases, indicates the transition's multipolar character. For example, weighting the transition density by the position vector yields the transition dipole moment ($\boldsymbol{\mu} = \int \mathbf{r} \rho^{g \to e}(\mathbf{r}) d\mathbf{r}$), which dictates the far-field optical response. The full TDC is given by the 3D spatial integral
\begin{equation}
    J_{\mathrm{TDC}} \approx \frac{1}{4\pi\varepsilon(t)}\iint \frac{\left(\rho^{g\to e}_L(\mathbf r)\right)^*\,\rho^{g\to e}_R(\mathbf r')}{|\mathbf r-\mathbf r'|} \,d\mathbf r\,d\mathbf r'
    \label{eq:J_integral}
\end{equation}
for time-dependent permittivity $\varepsilon(t)$.
Formally, the electron transition density can be complex, and the conjugate ensures a real coupling strength. In our numerical implementation, the 3D continuous integral is evaluated following the Transition Density Cube (TDC) method introduced by Krueger et al.~\cite{krueger1998}, where space is discretised on a 3D Cartesian grid of voxel volume $\Delta V$. Each grid voxel $i$ carries an effective transition charge $q_i = \rho^{g\to e}(\mathbf{r}_i)\Delta V$, yielding
\begin{equation}
    J_{\mathrm{TDC}} \approx \frac{1}{4\pi\varepsilon(t)} \sum_{i\in L}\sum_{j\in R}\frac{q_i q_j}{|\mathbf r_i-\mathbf r_j|},
\end{equation}
as distinct from transition atomic charge representations (e.g., the TrESP method of Madjet, Abdurahman and Renger~\cite{madjet2006tresp}). For comparison, we also evaluate the far-field PDA, which collapses these densities into two idealised point dipoles $\boldsymbol{\mu}_m$ separated by vector $\mathbf{R}$ given by
\begin{equation} \label{eq:J_PD}
    \resizebox{.98\columnwidth}{!}{$\displaystyle
    J_{\mathrm{PDA}} = \frac{1}{4\pi\varepsilon(t)} \left( \frac{\boldsymbol{\mu}_L \cdot \boldsymbol{\mu}_R}{|\mathbf{R}|^3} - \frac{3(\boldsymbol{\mu}_L \cdot \mathbf{R})(\boldsymbol{\mu}_R \cdot \mathbf{R})}{|\mathbf{R}|^5} \right).$}
\end{equation}

The Coulomb integral is screened by a time-dependent effective permittivity $\varepsilon(t)$ that tracks the solvent response. The aqueous environment relaxes with a Debye time $\tau_D \approx 8.3$\,ps; because this timescale is orders of magnitude longer than the sub-picosecond timescale of excitation transfer, the solvent cannot undergo nuclear equilibration during the transfer. The permittivity is therefore initially governed by the optical dielectric constant ($\varepsilon_{\rm opt} \approx 1.78$) and only later approaches the static limit ($\varepsilon_{\rm eq} \approx 78$), as described by
\begin{equation}
    \frac{1}{\varepsilon(t)} = \frac{1}{\varepsilon_{\rm eq}} + \left(\frac{1}{\varepsilon_{\rm opt}} - \frac{1}{\varepsilon_{\rm eq}}\right)e^{-t/\tau_D}.
\end{equation}

Within the sub-picosecond window ($t < 1$\,ps) over which electronic dephasing and population localisation occur in our dynamical simulations, the inverse dielectric screening factor $1/\varepsilon(t)$ changes only marginally, from $1/\varepsilon_{\rm opt} \approx 0.562$ at $t = 0$ to $\approx 0.501$ at $t = 1$\,ps (an $\sim 11\%$ decrease). Consequently, this small variation means that during the primary dephasing event ($T_2^* = 60$\,fs) and initial site localisation, the excitonic interaction is predominantly governed by the optical-limit coupling $J(0)$. The slow, macroscopic solvent relaxation toward the fully screened static limit ($\varepsilon_{\rm eq} \approx 78$, where $1/\varepsilon \approx 0.013$) unfolds on the much longer $\tau_D \approx 8.3$\,ps timescale, well after coherent phase relationships have already been damped by environmental fluctuations.

Assuming a structurally identical homodimer where the ground to first excited state excitation energies are identical for each monomer ($E_L = E_R = E$), the excited state Hamiltonian in the site basis is
\begin{equation}
    \hat H_{\text{site}}(t)=\begin{pmatrix} E & J(t) \\ J(t) & E \end{pmatrix}.
\end{equation}
Diagonalising $\hat H_{\text{site}}(t)$ yields the adiabatic (exciton) basis, where eigenstates are the symmetric (bright) and antisymmetric (dark) combinations such as $\ket{\pm} = \frac{1}{\sqrt{2}}(\ket{1} \pm \ket{2})$. Photoexcitation is treated as a vertical transition directly into the delocalised bright state $\ket{+}$, and its energy separation is the Davydov splitting, given by $\Delta E(t) = 2|J(t)|$. The dynamics are seeded with the optical-limit coupling at the instant of absorption, $J(0)$. Over several picoseconds, the coupling $J(t)$ relaxes as the solvent environment reorganises from the initial optical dielectric response toward the static equilibrium limit, resulting in a gradual damping of the excitonic interaction.

Because the physical interactions between the system and its aqueous environment are local, we define the master equation (ME) with a Hermitian jump operator $\hat{L} = \sqrt{\hbar \gamma / 2} \hat{\sigma}_z$ in the site basis and the pure dephasing rate $\gamma = 1/T_2^*$ ($\hat{\sigma}_z$ for the Pauli-$z$ operator and $\hbar$ for the reduced Planck constant). For our simulations, a homogeneous pure-dephasing time of $T_2^* = 60$\,fs was used, the rate at which environmental fluctuations randomise the relative phase of the two site states and damp the off-diagonal coherence $\rho_{12}$, as distinct from the population-relaxation time $T_1$. While the rigid $\beta$-barrel of fluorescent proteins can partially shield chromophores from solvent fluctuations and extend coherence relative to unbound organic dyes \cite{kim2019venusa206}, this conservatively fast $60$\,fs limit was chosen to align with empirical measurements of rapidly decaying electronic quantum coherence in solvated pigment-protein complexes at ambient temperatures \cite{duan2017nature}. {We stress that $T_2^*$ enters as a phenomenological, order-of-magnitude parameter transferred from these analogous systems: no $T_2^*$ has been measured for the Venus dimer, and, importantly, it is deliberately \emph{not} extracted from our classical thermal trajectory. Homogeneous pure dephasing is a property of the \emph{quantum} bath correlation function (it obeys detailed balance and is dominated by high-frequency intramolecular modes with $\hbar\omega \gg k_{\rm B}T$), and cannot be obtained rigorously from classical molecular dynamics, which samples equilibrium configurational statistics rather than the quantum bath response. What the classical ensemble \emph{does} legitimately supply here is the complementary \emph{inhomogeneous} broadening from the static spread of the coupling over thermally sampled geometries (the $\pm 1.6~\mathrm{cm^{-1}}$ distribution of $J$ in Fig.~\ref{fig:tandem}); site-energy disorder is not sampled in the present $1$\,ns ensemble. Crucially, our argument never requires the precise value of $T_2^*$: it enters only as an upper bound on the dephasing timescale, so that the central claim, that exciton formation and the Davydov splitting are imprinted faster than the environment can dephase or dielectrically relax, holds for any $T_2^*$ short compared with the $\sim 8.3$\,ps Debye time. We confirm this by sweeping $T_2^*$ over $20$--$200$\,fs: the resulting coherence-decay times ($\sim 0.02$--$0.2$\,ps) all remain orders of magnitude below the $\sim 8.3$\,ps Debye time, so $T_2^*$ sets only the rate of localisation in Fig.~\ref{fig:ExcitonDynamics} while the separation of timescales itself is preserved throughout.} Our choice of jump operator effectively ``measures'' the excitation site, driving the state towards the local site basis, which emerges as the decoherence-resistant pointer basis ($\ket{1}$ and $\ket{2}$) since physical interactions are local. The system's ensemble-averaged density matrix $\hat{\rho}(t)$ in the $\{\ket{1},\ket{2}\}$ basis evolves via the Lindblad equation \cite{manzano2020short}, given by
\begin{equation}
    \frac{d\hat{\rho}}{dt}=-\frac{i}{\hbar}[\hat H_{\text{site}}(t),\hat\rho] + \hat{L} \hat\rho \hat{L} - \frac{1}{2} \{\hat{L}^2, \hat\rho\}\,.
    \label{eq:ME01}
\end{equation}

Building on the Lindblad model, we extend the analysis by simulating individual pure-state trajectories in addition to the ensemble density operator \cite{abrahams2025excitonic}. While the ME provides a deterministic description of the average behaviour of a large ensemble of dimers, it cannot resolve the discrete, incoherent events experienced by a single system. To capture these dynamics, we use a stochastic Schrödinger equation (SSE), which describes the evolution of a single pure-state trajectory $|\psi\rangle$ subject to continuous environmental monitoring. The density operator is recovered by taking the stochastic average $\mathbb{E}$ over many trajectories, $\hat{\rho} = \mathbb{E}[|\psi\rangle\langle\psi|]$. Then, the SSE is given by
\begin{multline}
    |d\psi\rangle = \left( -\frac{i}{\hbar}\hat{H}_{\text{site}} - \frac{1}{2}(\hat{L} - \langle \hat{L} \rangle)^2 \right) |\psi\rangle dt \\
    + (\hat{L} - \langle \hat{L} \rangle) |\psi\rangle dW \, ,
\end{multline}
where $dW$ is an It\^{o} Wiener process~\cite{petruccione}.

We visualise the individual pure-state trajectories (\textbf{Figures~\ref{fig:ExcitonDynamics}(b) and (c)}) by mapping the system to the Bloch vector $\mathbf{r} = (\langle\hat\sigma_x\rangle, \langle\hat\sigma_y\rangle, \langle\hat\sigma_z\rangle)$ (\textbf{Figure~\ref{fig:ExcitonDynamics}(a)}), where $\langle\hat\sigma_x\rangle = 2\text{Re}(\rho_{12})$, $\langle\hat\sigma_y\rangle = 2\text{Im}(\rho_{12})$, and $\langle\hat\sigma_z\rangle = \rho_{11} - \rho_{22}$. All the numerical simulations are initialised in the delocalised bright state $\ket{+}$, corresponding to a vector on the equator of the Bloch sphere. Individual trajectories exhibit incoherent hopping between sites in the bright-dark basis. These stochastic paths represent pure-state evolution restricted to the surface of the sphere as shown in \textbf{Figure~\ref{fig:ExcitonDynamics}(a)}. Conversely, the ensemble-averaged dynamics governed by the ME in Eq.~(\ref{eq:ME01}) describe the rapid, smooth decay of off-diagonal coherence, leading to a statistical mixture over both sites and evolution toward the centre of the Bloch sphere, shown as the solid blue line in \textbf{Figure~\ref{fig:ExcitonDynamics}(a)}.

\begin{figure*}[t]
  \centering
  \includegraphics[width=0.31\textwidth]{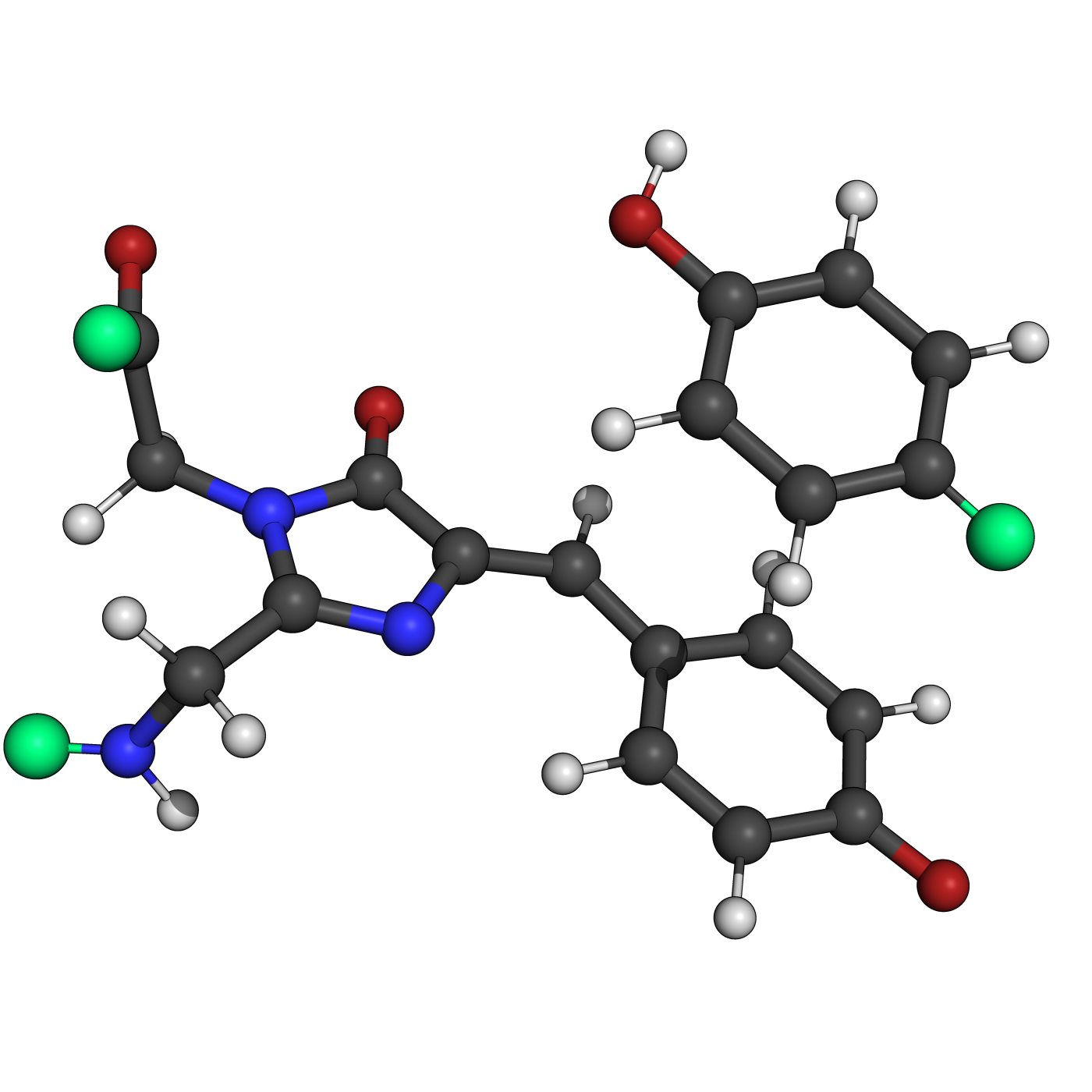}\hfill
  \includegraphics[width=0.31\textwidth]{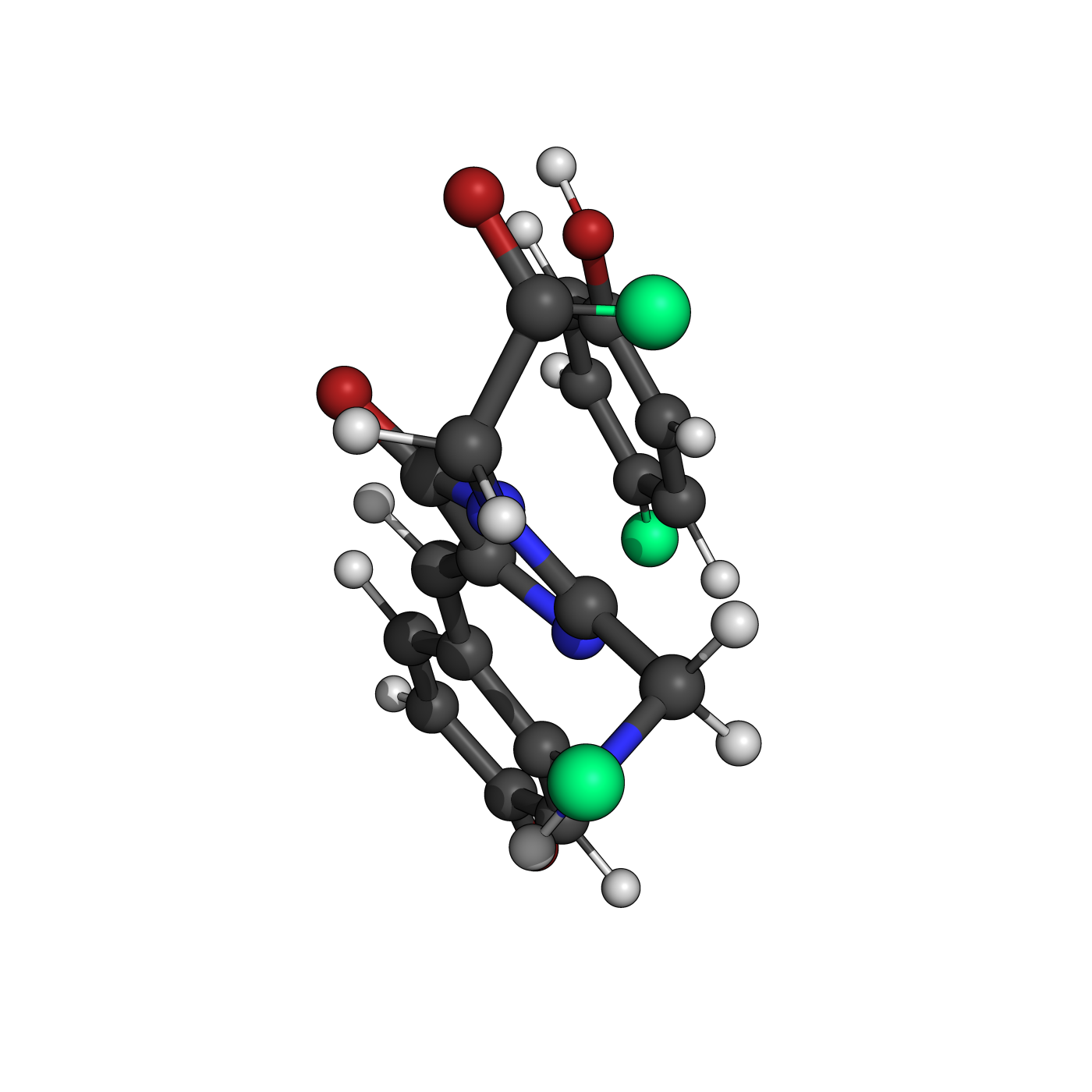}\hfill
  \includegraphics[width=0.31\textwidth]{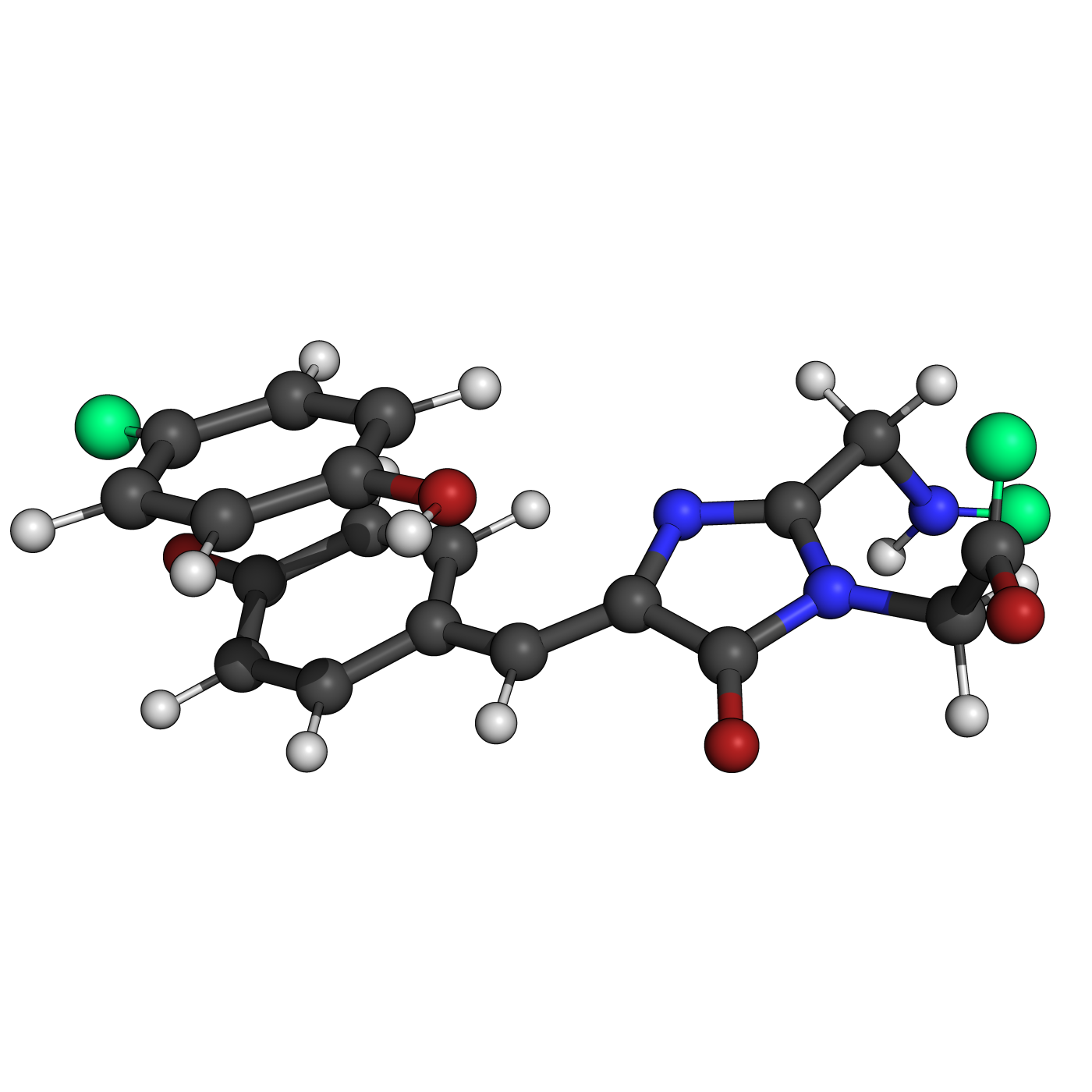}
  \caption{The $44$-atom QM region used for electronic-structure calculations, drawn from the production geometry at three orientations. Carbon grey, oxygen red, nitrogen blue, hydrogen white; the three link hydrogens are green and enlarged. The anionic CR2 chromophore ($31$ atoms including two backbone link-H caps) comprises the imidazolinone at centre and the phenolate ring at lower right; the $\pi$-stacked Tyr203 phenol ($13$ atoms including one link-H cap at the $\mathrm{C_\beta-C_\gamma}$ cut) retains its hydroxyl. Net charge $-1$, stoichiometry $\mathrm{C_{19}H_{18}N_3O_4}$.}
  \label{fig:qmregion}
\end{figure*}

\section{Excited-State Structure and Excitonic Coupling}

\begin{table}[t]
\centering
\caption{Photon wavelength ($\lambda$) and oscillator strength ($f$) of the vertical bright $\pi\pi^{*}$ excitation of the anionic Venus chromophore in the same frozen $44$-atom QM geometry, computed with the protein/solvent MM point-charge embedding (QM/MM) and after removal of all point charges (gas phase). TDDFT yields a dominant bright $\pi\pi^{*}$ root but blueshifts it; STEOM-CCSD, adopted here as the method of record, agrees with the measured absorption in the embedded model. Comparison between QM/MM and gas-phase values demonstrates an environmental solvatochromic blueshift of $+386~\mathrm{cm^{-1}}$ (TDDFT) and $+722~\mathrm{cm^{-1}}$ (STEOM-CCSD) induced by the protein electrostatic field.}
\label{tab:roots}
\vspace{4pt}
\resizebox{\columnwidth}{!}{%
\begin{tabular}{lcccc}
\toprule
& \multicolumn{2}{c}{QM/MM} & \multicolumn{2}{c}{Gas Phase} \\
\cmidrule(lr){2-3}\cmidrule(lr){4-5}
Method & $\lambda$ (nm) & $f$ & $\lambda$ (nm) & $f$ \\
\midrule
TDDFT, $\omega$B97X-D3      & 360.92 & 1.033 & 366.02 & 1.262 \\
DLPNO-STEOM-CCSD            & 523.90 & 0.885 & 544.51\footnotemark[1] & 0.916 \\
Experiment (absorption)     & 514.46 & n/a   & -- & -- \\
\bottomrule
\end{tabular}}\par\vspace{2pt}
{\footnotesize \footnotemark[1]In the gas phase, the isolated anion lacks stabilizing dielectric embedding; the vacuum calculation is included solely for direct frozen-geometry comparison to quantify the electrostatic environment shift.}
\end{table}

\begin{table}[t]
\centering
\caption{Method calibration on the gas-phase (neutral) chromophore, def2-SVP. Bare EOM-CCSD predicts the excitation at too short a wavelength; the non-iterative triples correction of EOM-CCSD(fT) redshifts it from $333.29$ to $376.85~\mathrm{nm}$ (a shift of $43.56~\mathrm{nm}$), close to the DLPNO-STEOM-CCSD value of $371.21~\mathrm{nm}$, and ADC(2) assigns ${\sim}12\%$ doubles (2p2h) character to the transition. STEOM-CCSD thus reproduces the triples-corrected benchmark, confirming that it captures the double-excitation physics that single-excitation TDDFT cannot represent.}
\label{tab:ladder}
\begin{tabular}{lc}
\toprule
Method & $\lambda$ (nm) \\
\midrule
EOM-CCSD (bare)           & 333.29 \\
EOM-CCSD(fT) (+triples)   & 376.85 \\
DLPNO-STEOM-CCSD          & 371.21 \\
\bottomrule
\end{tabular}
\end{table}

\begin{figure*}[t]
  \centering
  \begin{subfigure}[b]{0.48\textwidth}
    \centering
    \includegraphics[width=\linewidth]{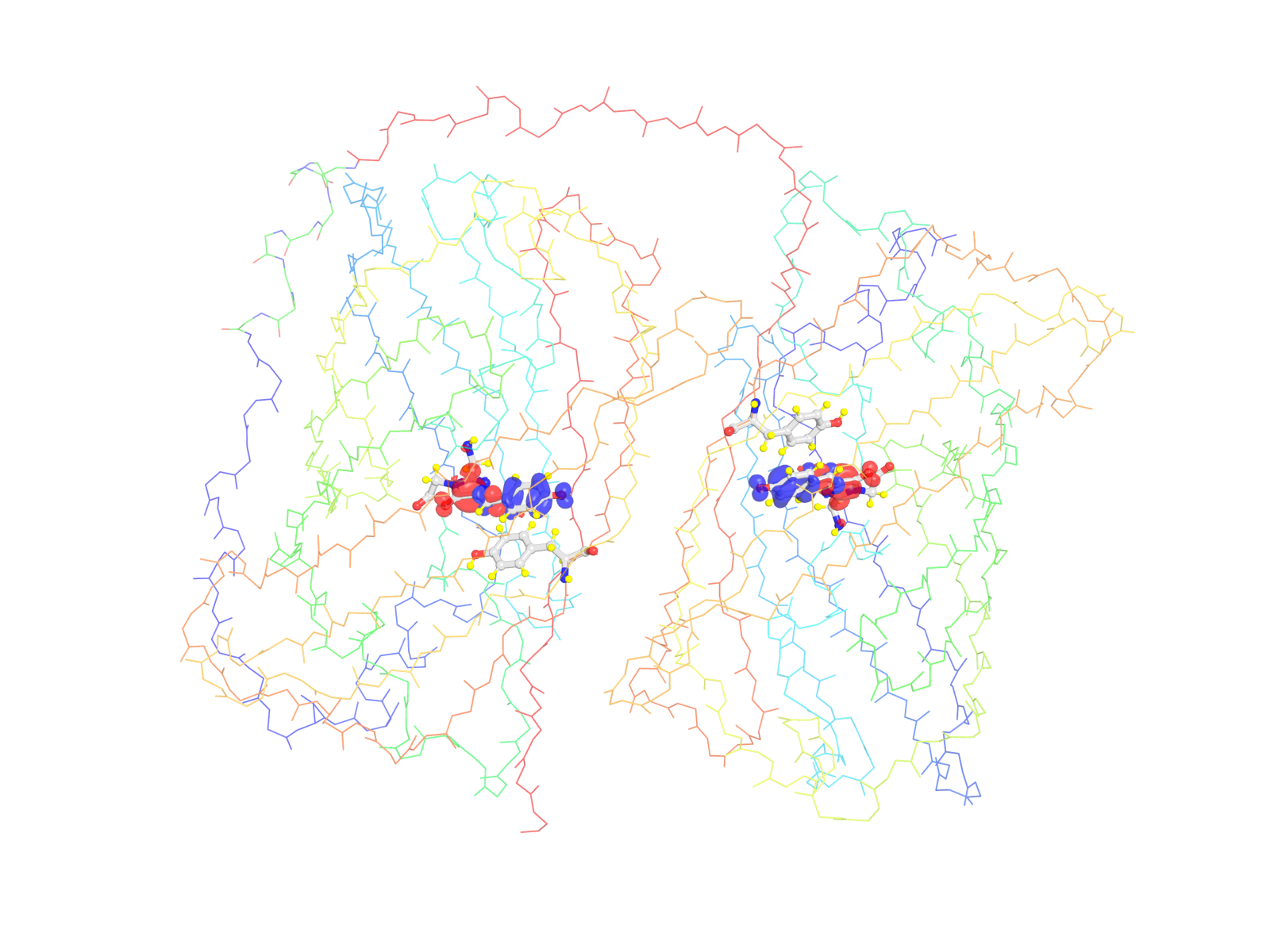}
    \caption{}
    \label{fig:density_dimer}
  \end{subfigure}\hfill
  \begin{subfigure}[b]{0.48\textwidth}
    \centering
    \includegraphics[width=\linewidth]{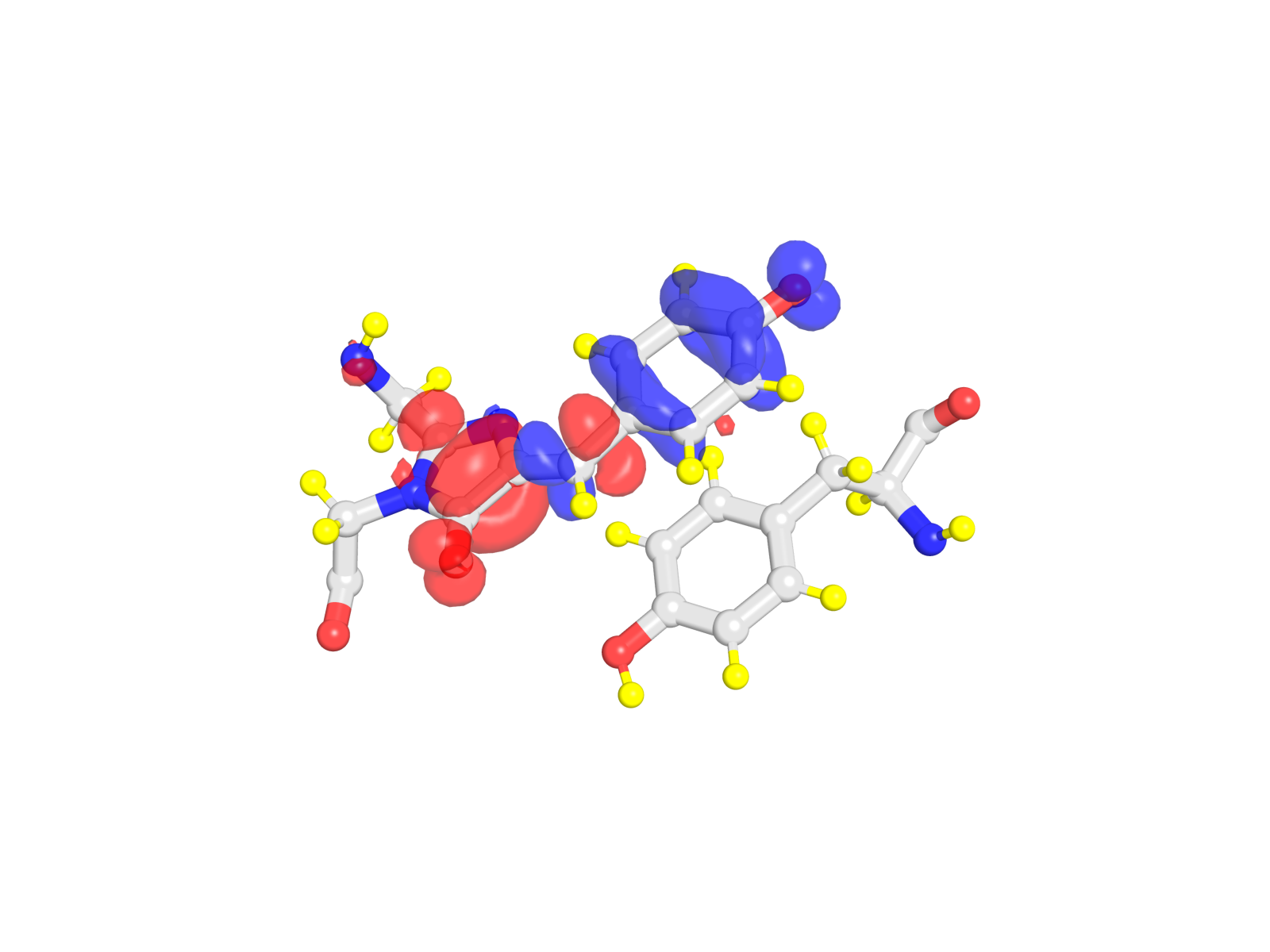}
    \caption{}
    \label{fig:density_qm}
  \end{subfigure}
  \caption{Visualisation of the electron transition density $\rho^{g\to e}_m(\mathbf r)$ in Eq.~(\ref{eq:transition_density})  between the ground and first excited state in Table \ref{tab:roots}. The red and blue isosurfaces represent the positive and negative phases of the transition density, mapped onto the dimer frame using PyMOL structural alignment. (a) The full dimer system partitioned into QM and MM domains. The MM embedding environment, treated as point charges, is rendered as a protein backbone with a rainbow colourmap to give a sense of depth perception. (b) A detailed view of the $44$-atom STEOM QM region (indicated with sticks and spheres) on the right monomer, comprising the deprotonated CR2 chromophore and the $\pi$-stacked Tyr203 phenol on which the correlated transition density is evaluated. PyMOL \cite{delano2002pymol} generation scripts are available on GitHub.}
  \label{fig:density_combined}
\end{figure*}

\begin{figure*}[t]
  \centering
  \begin{subfigure}[b]{0.48\textwidth}
    \centering
    \includegraphics[width=\linewidth]{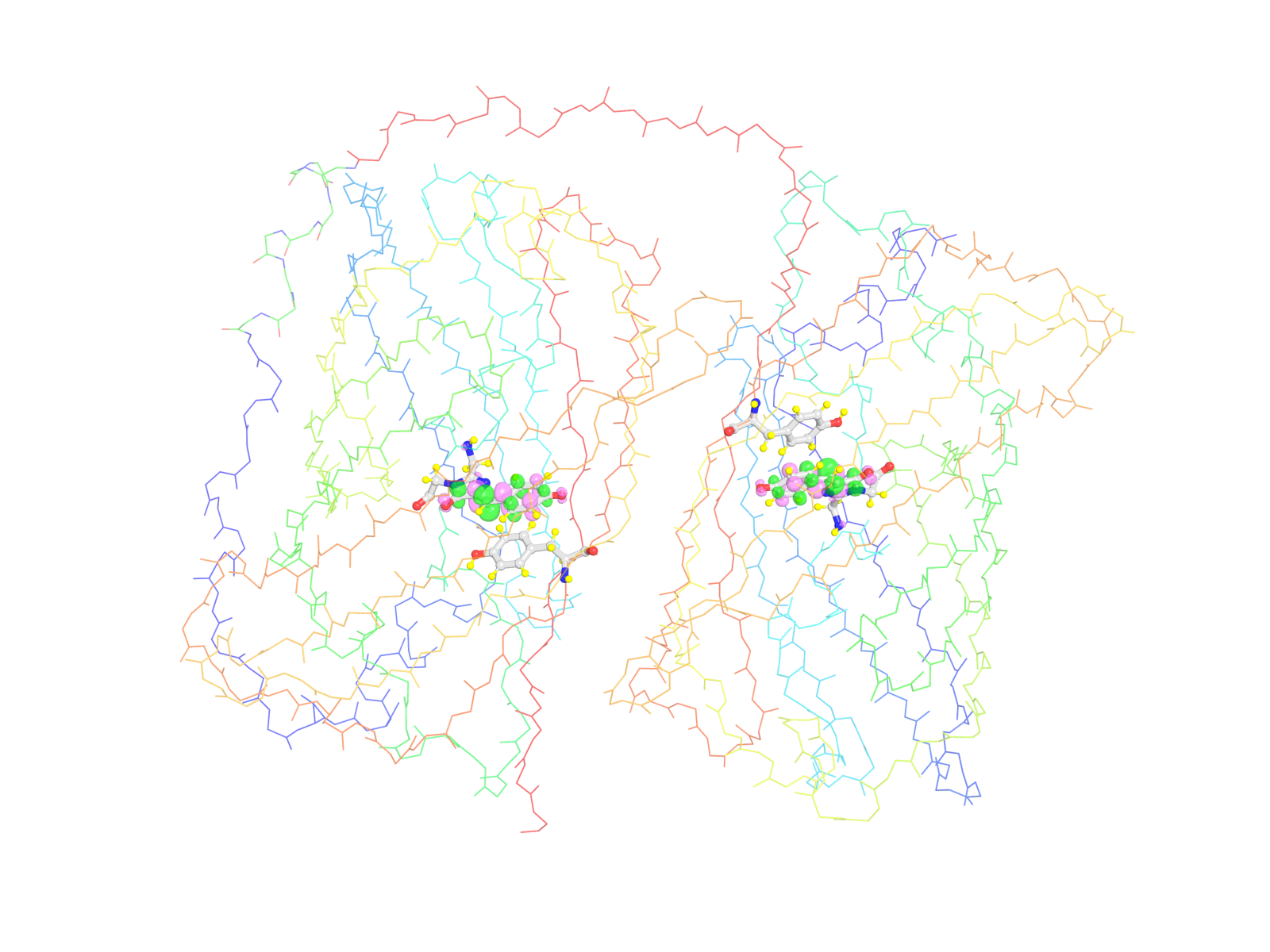}
    \caption{}
    \label{fig:electron_dimer}
  \end{subfigure}\hfill
  \begin{subfigure}[b]{0.48\textwidth}
    \centering
    \includegraphics[width=\linewidth]{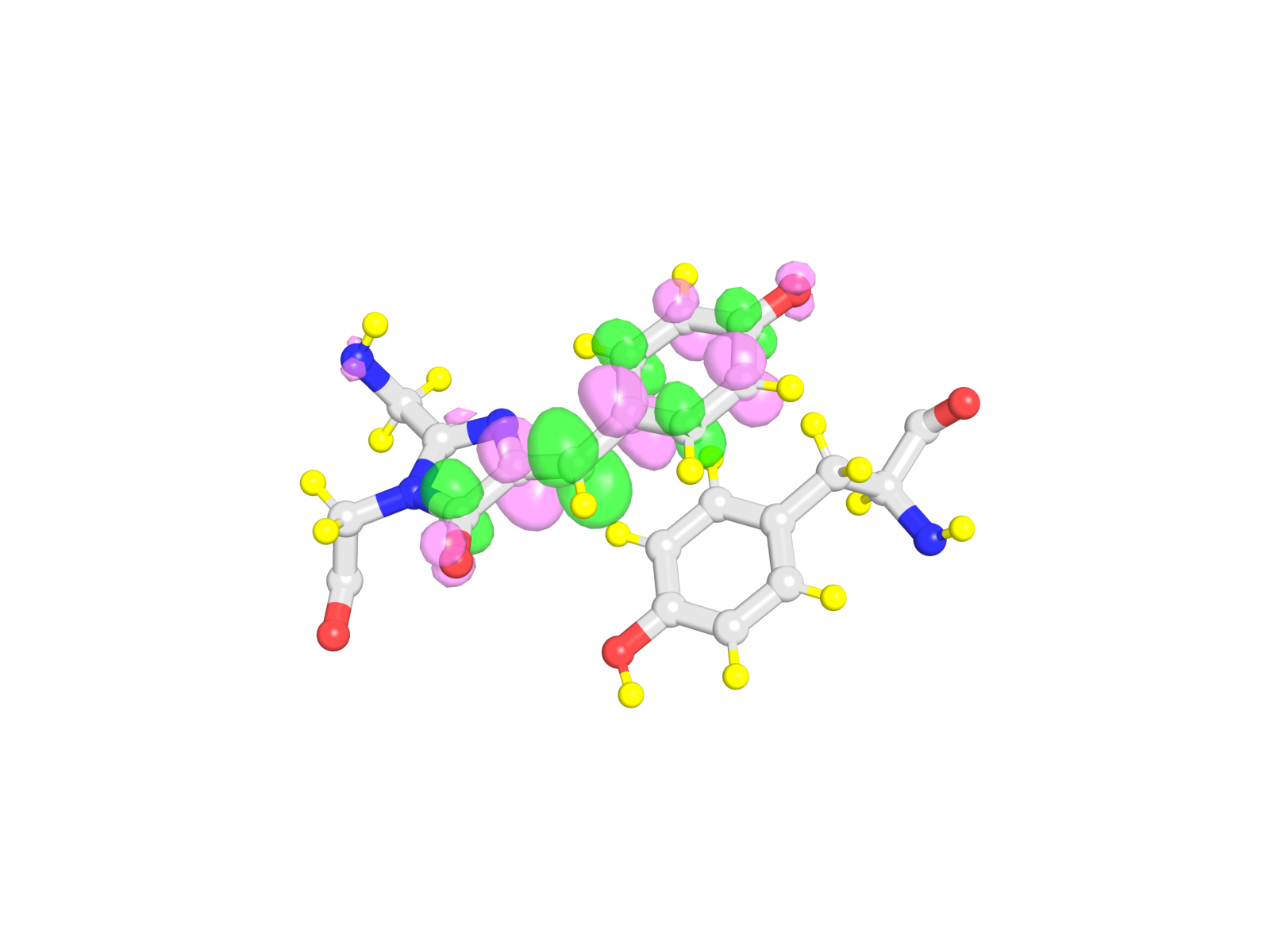}
    \caption{}
    \label{fig:electron_qm}
  \end{subfigure}
  \caption{Visualisation of the electron density difference ($\rho^{e} - \rho^{g}$) between the ground and first excited state delocalised over the dimeric Venus geometry from Table \ref{tab:roots}. The green and violet isosurfaces represent electron gain and loss respectively. (a) The dimer frame illustrating the QM/MM partitioning, with the MM backbone rendered in the rainbow colourmap to give a sense of depth perception. (b) A detailed view of the $44$-atom STEOM QM region on the right monomer, comprising the deprotonated CR2 chromophore and the $\pi$-stacked Tyr203 phenol. The detailed PyMOL \cite{delano2002pymol} codes are available on GitHub.}
  \label{fig:electron_combined}
\end{figure*}

The chromophore possesses a single bright excitation in the visible: TDDFT ($\omega$B97X-D3, computed on the same $44$-atom QM region and point-charge field as the correlated calculation) returns a dominant bright $\pi\pi^{*}$ root ($f=1.03$), but as a single-excitation theory it blueshifts this state to $360.42~\mathrm{nm}$, roughly $154~\mathrm{nm}$ to the blue of experiment. We therefore adopt the correlated wavefunction result as the site energy: the converged DLPNO-STEOM-CCSD electric-dipole spectrum (ORCA \cite{neese2020orca,neese2022orca,nooijen1997steom,dutta2018ipeom,dutta2019eaeom,dutta2016dlpno}) places the in-protein bright state at $523.90~\mathrm{nm}$ ($f=0.885$), in good agreement with the experimental absorption maximum near $515~\mathrm{nm}$ (Table~\ref{tab:roots}). This assignment is validated on the tractable gas-phase (neutral) chromophore, where a non-iterative triples correction redshifts bare EOM-CCSD from $333.29$ to $376.85~\mathrm{nm}$, close to the STEOM value of $371.21~\mathrm{nm}$ (Table~\ref{tab:ladder}) \cite{manohar2008ft,epifanovsky2021qchem}. To illustrate the QM partitioning, \textbf{Figure~\ref{fig:qmregion}} details the $44$-atom QM region at three orthogonal orientations, showing the deprotonated CR2 core and $\pi$-stacked Tyr203 phenol. To physically characterise the transition, the STEOM electron transition density ($\rho^{g\to e}$) and the ground-to-excited state electron density difference ($\rho^{e} - \rho^{g}$) were mapped onto the dimer frame (\textbf{Figure~\ref{fig:density_combined}} and \textbf{Figure~\ref{fig:electron_combined}}, respectively).

The electron transition density in \textbf{Figure~\ref{fig:density_combined}}, which serves as the fundamental quantity for evaluating the Coulombic integral in Eq.~(\ref{eq:J_integral}), exhibits a complex multipolar spatial distribution in 3D that extends across the entire chromophore and adjacent residues. Note that the red and blue isosurfaces in \textbf{Figure~\ref{fig:density_combined}(b)} do not represent static charge gain or loss, but rather the spatial ``lobes'' of the electron transition probability. The spatial overlap of these lobes between the two chromophores determines the magnitude of $J_{\text{TDC}}$, and captures interactions that are ignored when the density is treated as a single point-dipole vector. Concurrently, the electron density difference in \textbf{Figure~\ref{fig:electron_combined}(b)} illustrates the charge reorganisation from the ground to the first excited state (state 1), with the deprotonated phenolic oxygen on the chromophore residue (CR2, anionic form) transferring charge to the rest of the chromophore. 

The STEOM transition densities were mapped onto the dimer frame (chromophore centroid separation $24.7~\text{\AA}$; inter-dipole angle $101^{\circ}$). This computed geometry is consistent with the two-photon polarization ratio measured for the Venus dimer, which independently constrains the relative orientation of the two chromophore transition dipoles \cite{cusick2026polarization}. At this geometry, the full transition-density coupling is in close agreement with the far-field point-dipole approximation:
\begin{itemize}
    \item \textbf{Point-Dipole Approximation:} $J_{\text{PDA}} = 27.6~\mathrm{cm^{-1}}$,
    \item \textbf{Transition Density Coupling:} $J_{\text{TDC}} = 32.82~\mathrm{cm^{-1}}$,
    \item \textbf{Near-field ratio:} $J_{\text{TDC}}/J_{\text{PDA}} \approx 1.19$.
\end{itemize}
\noindent For comparison, evaluating the same TDC pipeline on the TDDFT transition density gives $J_{\text{TDC}} = 27.7~\mathrm{cm^{-1}}$ versus $J_{\text{PDA}} = 25.0~\mathrm{cm^{-1}}$ ($J_{\text{TDC}}/J_{\text{PDA}} \approx 1.11$); the modest near-field enhancement over the PDA (${\sim}10$--$20\%$) is thus a robust, method-consistent feature.

The close agreement between the TDC calculation and the far-field point-dipole estimate reflects the well-separated nature of the two chromophores. At a centroid separation of $24.7~\text{\AA}$, the point-dipole approximation provides a reliable leading-order estimate of the coupling strength, with higher-order multipoles contributing a modest $\sim 10$--$20\%$ enhancement. This is in quantitative agreement with independent direct coupling benchmark calculations on the Venus crystal structure (e.g., Q-Chem direct coupling yielding $48.7~\mathrm{cm^{-1}}$ for TDC versus $43.2~\mathrm{cm^{-1}}$ for PDA, a $13\%$ difference). A Cartesian multipole decomposition reveals that the dipole--dipole term accounts for the dominant portion of the interaction, while dipole--quadrupole and higher multipolar terms provide the remaining $\sim 10$--$20\%$ correction.

To identify the origin of this enhancement, we decomposed the coupling into a primitive Cartesian multipole series about each chromophore centroid (companion analysis \texttt{multipole\_analysis.py}, evaluated on the transition density of the bright $\pi\pi^{*}$ state), evaluating the transition dipole, quadrupole and octupole moments and the corresponding inter-site interaction terms (dipole--dipole, dipole--quadrupole, quadrupole--dipole, dipole--octupole, quadrupole--quadrupole, octupole--dipole). The dipole--dipole term reproduces $J_{\text{PDA}}$ exactly by construction, but the truncated series converges only slowly at this geometry: summing all terms through octupole recovers just $\sim 28\%$ of the full $J_{\text{TDC}}$ (dipole alone $\sim 23\%$, through quadrupole $\sim 27\%$). The majority of the near-field enhancement therefore does not reside in a few leading multipoles but in the short-range \emph{Coulomb} coupling between the spatially extended transition densities, whose facing tails approach to $\sim 9~\mathrm{\AA}$, far closer than either the $\sim 15~\mathrm{\AA}$ nuclear closest approach or the $\sim 25~\mathrm{\AA}$ chromophore centroid separation, and which only the full transition-density integral captures. This is a purely electrostatic effect between non-overlapping densities, distinct from short-range Dexter exchange, which is negligible at these separations. This sharpens the broader lesson for the PDA: when transition densities are delocalised and their nearest approach is comparable to the molecular size, even a multipole-corrected dipole model is inadequate and the full density is required. The decomposition is validated by a self-test that reconstructs an exact brute-force coupling between two compact, well-separated model densities to better than $0.1\%$.

\begin{figure*}[t]
  \centering
  \begin{subfigure}[b]{0.32\textwidth}
    \centering
    \includegraphics[width=\linewidth]{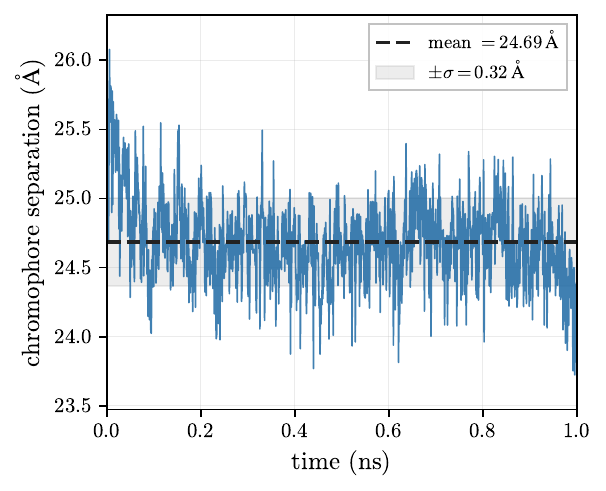}
    \caption{}
    \label{fig:tandem_sep}
  \end{subfigure}\hfill
  \begin{subfigure}[b]{0.32\textwidth}
    \centering
    \includegraphics[width=\linewidth]{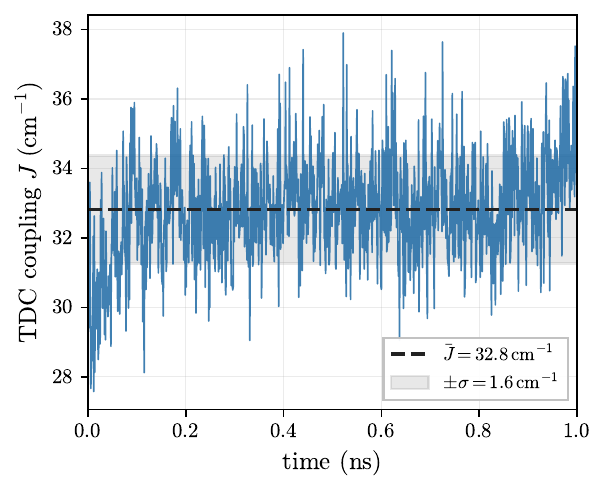}
    \caption{}
    \label{fig:tandem_coupling}
  \end{subfigure}\hfill
  \begin{subfigure}[b]{0.32\textwidth}
    \centering
    \includegraphics[width=\linewidth]{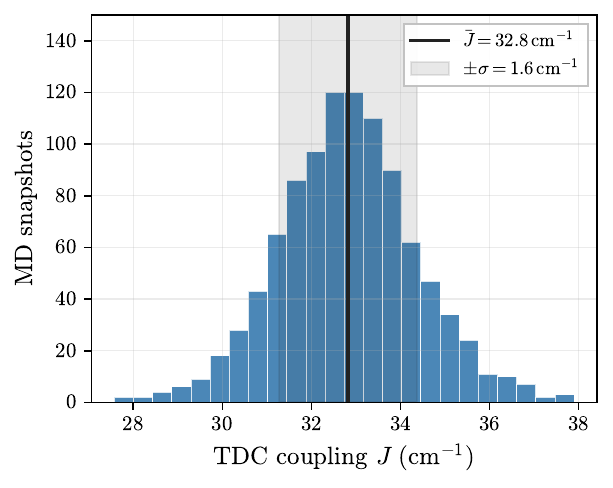}
    \caption{}
    \label{fig:tandem_hist}
  \end{subfigure}

  \caption{Dynamics of the VenusA206 tandem dimer ($1000$ frames saved every $1$\,ps over a $1$\,ns, $300$\,K unrestrained NVT trajectory; \texttt{run\_nvt.py}, \texttt{coupling\_dcd\_steom.py}). The two anionic CR2 residues use the production AMBER parameters, and every snapshot is evaluated with the complete cap-masked STEOM transition density ($259{,}277\times259{,}277$ mutual Coulomb sum, GPU, $\varepsilon_{\rm opt}=1.77$). (a) Chromophore-centroid separation versus time, $24.69\pm0.32~\mathrm{\AA}$ (mean dashed; shading is one sample standard deviation). (b) TDC coupling per snapshot, $J=32.8\pm1.6~\mathrm{cm^{-1}}$ across the $1000$-frame trajectory (sample mean dashed; shading is one sample standard deviation). (c) Distribution of all $1000$ snapshot values ($\bar{J} = 32.8~\mathrm{cm^{-1}}$, $\sigma = 1.6~\mathrm{cm^{-1}}$); the spread is the sample standard deviation.}
  \label{fig:tandem}
\end{figure*}

Because protein conformational dynamics modulate the inter-chromophore geometry and hence $J$, we characterised the coupling over a room-temperature trajectory of the actual VenusA206 tandem-dimer construct (see Methods, \textbf{Figure~\ref{fig:tandem}}). The covalently tethered tandem holds the A206 interface docked for the full $1~\mathrm{ns}$: the chromophore separation is $24.69 \pm 0.32~\mathrm{\AA}$ (range $23.72$--$26.08~\mathrm{\AA}$), and the full-grid transition-density coupling is $J = 32.8 \pm 1.6~\mathrm{cm^{-1}}$ across the $1000$-frame trajectory (Davydov splitting $2|J| = 65.6 \pm 3.1~\mathrm{cm^{-1}}$). The corresponding ensemble point-dipole result is $J_{\mathrm{PDA}} = 27.6 \pm 1.3~\mathrm{cm^{-1}}$, confirming a robust $19\%$ near-field enhancement. The covalent linker of the tandem construct sustains the docked geometry throughout the dynamics, confirming that intermediate coupling is an intrinsic feature of the tethered dimer.

Nguyen et al.\ analysed the $\mathrm{dVenus\mbox{-}TDX}-\mathrm{dVenus\mbox{-}TD}$ difference-molar-ellipticity spectrum with three Lorentzians \cite{nguyen2025anomalous}. Their constrained sensitivity fit (Table~S3) places the excitonic pair at $514.07$ and $521.07~\mathrm{nm}$, giving $U=130.79~\mathrm{cm^{-1}}$ and $2U=261.58~\mathrm{cm^{-1}}$, with a third band at $481.37~\mathrm{nm}$; the unconstrained fit shown in their main Fig.~4(d) instead gives $511.8$ and $521.7~\mathrm{nm}$, or $U\approx185.4~\mathrm{cm^{-1}}$. These are alternative fit conventions rather than the endpoints of a statistical confidence interval.

\section{Computational Methods}

The quantum-classical hybrid workflow is automated via a single Python orchestrator, \texttt{reproduce\_paper.py}, that integrates structural preparation, classical relaxation, QM/MM electronic-structure calculations (TDDFT, DLPNO-STEOM-CCSD, and the EOM-CCSD(fT)/ADC(2) validation), GPU-accelerated TDC analysis, and the thermal ensemble, spanning the ORCA and Q-Chem back-ends behind a common configuration (random seed, dielectric, ensemble size, per-stage caching). Algorithm \ref{alg:pipeline} serves as a concise, reproducible map of this orchestration, linking each manuscript quantity to the routine that produces it in the released code, rather than introducing new methodology; readers interested only in the physical results may skip it.

The simulation of the dimeric Venus (PDB: 1MYW \cite{berman2000protein}) proceeded through the following stages.

\begin{enumerate}
    \item \textbf{Structural Preparation and Relaxation:} Geometries for both the isolated monomer and the full interacting dimer were prepared and solvated using PDBFixer \cite{openmm2017} and OpenMM \cite{openmm2017}. Classical energy minimisation of both systems was performed to generate a relaxed AMBER14 \cite{amber14_manual} force field environment. The relaxed monomer geometry was extracted to define the Quantum Mechanical (QM) region for subsequent electronic structure calculations, while the relaxed dimer geometry was retained exclusively as the 3D spatial frame for the later coupling analysis. To partition the system for QM/MM embedding it is necessary to sever the protein backbone; to avoid introducing artificial chemical instability, these severed bonds were capped with hydrogen atoms.

    \item \textbf{TDDFT reference (ORCA):} As a single-excitation reference, excited-state roots were computed via TDDFT (range-separated hybrid $\omega$B97X-D3/6-311G** \cite{mardirossian2017}, Tamm--Dancoff approximation) in ORCA 6.1 \cite{neese2020orca,neese2022orca} on the \emph{same} $44$-atom QM region and MM point-charge field as the DLPNO-STEOM-CCSD calculation, so that the single-excitation and correlated site energies are compared on an identical structure and embedding.

    \item \textbf{Correlated site energy and transition density (ORCA):} Because the anionic chromophore carries appreciable double-excitation character, the site energy and the transition density used for the coupling were computed with DLPNO-STEOM-CCSD \cite{nooijen1997steom,dutta2016dlpno,dutta2018ipeom,dutta2019eaeom,riplinger2013dlpno} as implemented in ORCA 6.1 \cite{neese2020orca,neese2022orca}, using the def2-SVPD basis \cite{weigend2005def2,rappoport2010svpd} with the RIJCOSX approximation, TightSCF/TightPNO thresholds and an automatically constructed auxiliary basis. STEOM-CCSD reaches the excited states through a second similarity transformation of the coupled-cluster Hamiltonian $\bar H = e^{-\hat T}\hat H\,e^{\hat T}$, forming $\tilde H = \{e^{-\hat S}\bar H\,e^{\hat S}\}$ with $\hat S = \hat S^{\mathrm{IP}} + \hat S^{\mathrm{EA}}$ built from ionization- and electron-attachment-EOM-CCSD amplitudes; diagonalising $\tilde H$ over single excitations folds in the double- and partial-triple-excitation (2p2h and factorisable 3p3h) character while avoiding the full connected triples $\hat T_3$. To keep the correlated calculation tractable while retaining the residue that controls the in-protein shift, Tyr203 was included because its $\pi$--$\pi$ stacking interaction with the conjugated CR2 chromophore directly perturbs the chromophore's electronic environment and site energy; the resulting QM region comprised the anionic chromophore and Tyr203 ($44$ atoms, hydrogen-capped, net charge $-1$), embedded in the same MM point-charge field. The active space was fixed deterministically through the DLPNO-STEOM double-excitation filter so that the reported bright state is reproducible run-to-run. The STEOM natural transition orbitals define the transition density $\rho^{g\to e}$, scaled to the STEOM oscillator strength. This correlated single-point was evaluated at the classically energy-minimised (local-minimum) geometry for numerical stability and run-to-run reproducibility of the reference density; the definitive inter-chromophore coupling is then obtained by evaluating this density over the thermally sampled tandem ensemble (below) rather than from any single geometry.

    \item \textbf{Method validation (Q-Chem):} The reliability of STEOM-CCSD for this transition was benchmarked on the tractable gas-phase (neutral) chromophore against canonical EOM-CCSD, its non-iterative triples correction EOM-CCSD(fT) \cite{manohar2008ft}, and ADC(2), all in Q-Chem \cite{epifanovsky2021qchem} (Table~\ref{tab:ladder}).

    \item \textbf{Transition Density Coupling (TDC):} The inter-chromophore coupling $J$ was evaluated by reconstructing the full dimeric interaction from the monomeric data. The transition density was duplicated and spatially mapped onto the two chromophore sites of the classically relaxed dimer geometry by a rigid (Kabsch) fit of the shared CR2 atoms, ensuring the STEOM density is placed in the same coordinate frame from which the dimer chains were built; direct numerical integration was then performed across the aligned grids. The density was represented on a uniform $200\times200\times200$ Cartesian grid with a spacing of $\approx 0.19~\mathrm{\AA}$ (voxel volume $\approx 7\times10^{-3}~\mathrm{\AA^3}$); retaining only voxels with $|\rho^{g\to e}| > 10^{-7}$ leaves $\sim 3\times10^{5}$ significant grid points that enter the Coulomb double sum, which is accumulated on the GPU (Numba-CUDA or PyOpenCL \cite{lam2015numba,klockner2012pycuda}). We note that such couplings can alternatively be evaluated analytically from atom-centred transition charges \cite{hsu2009}; the dense-grid integral used here converges to the same value and additionally supplies the volumetric densities of Figures~\ref{fig:density_combined}--\ref{fig:electron_combined}. Numerical results were subsequently screened by the optical dielectric constant ($\varepsilon_{\rm opt} \approx 1.78$) to account for the sub-picosecond timescale of exciton formation relative to bulk solvent relaxation.
\end{enumerate}

{To sample the room-temperature conformational dynamics that modulate the inter-chromophore geometry and hence $J$, we built the VenusA206 tandem dimer (VenusA206-TD): two Venus $\beta$-barrels covalently joined by the flexible ${\sim}33$-residue inter-domain linker used in the tandem constructs, with the barrels docked at the physiological A206 interface (residues clustered around Ala206, Leu207, Leu221 and Phe223; released tool \texttt{build\_\allowbreak tandem\_\allowbreak dimer.py}). The two CR2 sites received the complete bonded, nonbonded, RESP-charge, exclusion, and $1$--$4$ terms from the production anionic CR2 AMBER topology rather than generic fallback parameters. We then ran $500{,}000$ unrestrained OpenMM NVT steps at $300$\,K with a $2$\,fs timestep ($1$\,ns total), saving the protein and chromophores every $500$ steps to give $1000$ snapshots at $1$\,ps intervals (\texttt{run\_nvt.py}). Periodic imaging was removed by whole-residue lattice translations without coordinate fitting or deformation. For every frame, each copy of the cap-masked STEOM transition density was placed by a rigid fit of the $19$ shared CR2 heavy atoms and the mutual Coulomb integral was accumulated on an RTX 4080 over all $259{,}277^2$ density-point pairs in FP32 (\texttt{coupling\_dcd\_steom.py}); full-grid FP64 checks on frames 1, 500, and 1000 differed by only $0.011$--$0.014~\mathrm{cm^{-1}}$. The resulting separation and coupling distributions are reported with the Results (Fig.~\ref{fig:tandem}). Animations of the NVT trajectories are available at \url{https://www.youtube.com/playlist?list=PLbopStXxFpaQ}.}

\section{Summary}

We have characterised the excitonic coupling and energy-transfer dynamics in the Venus dimer using a hybrid QM/MM transition-density approach within an open-quantum-system model. Using a correlated (DLPNO-STEOM-CCSD) site energy and transition density (validated against an explicit triples benchmark) and simulating the actual VenusA206 tandem-dimer construct, we find that the covalent inter-domain linker holds the A206 interface docked throughout a nanosecond of dynamics, yielding $J = 32.8 \pm 1.6~\mathrm{cm^{-1}}$ across $1000$ snapshots and a Davydov splitting of $2|J| = 65.6 \pm 3.1~\mathrm{cm^{-1}}$, firmly within the intermediate excitonic regime. Distributed transition density calculations reveal that the point-dipole approximation ($J_{\mathrm{PDA}} = 27.6 \pm 1.3~\mathrm{cm^{-1}}$) provides a reliable leading-order description at this $\sim 24.7$~\text{\AA}\ separation, with higher multipoles contributing a modest $19\%$ near-field enhancement.

Our analysis validates the role of non-equilibrium dielectric screening in protein environments. Because the excitonic interaction timescale is significantly faster than the macroscopic Debye relaxation time of bulk water ($\sim 8.3$\,ps), the excitonic coupling can operate in the optical dielectric limit. This inertial ``freezing'' of the solvent protects the initial coupling from overwhelming equilibrium screening with $\varepsilon_{\rm eq}$. The stochastic trajectories illustrate that this coherence is transient and environmental interactions drive the system into a localised pointer basis on a sub-$100$\,fs timescale. Over the sub-picosecond timescale ($< 1$\,ps) of coherent exciton transfer, $1/\varepsilon(t)$ varies by only $\sim 11\%$, leaving the coupling effectively governed by $J(0)$. This temporal separation reconciles the Davydov splitting imprinted at absorption with rapid localisation and subsequent screening within picosecond timescales. The argument is not specific to Venus: because Debye relaxation of aqueous environments universally lies in the picosecond range while optical absorption and dephasing are far faster, the same optical-limit imprinting of the coupling is expected in other water-exposed multichromophore complexes, with the $\beta$-barrel here additionally suppressing the fluctuation amplitude, not the underlying separation of timescales. Although Fig.~\ref{fig:ExcitonDynamics} accounts well for the single-photon excited-subspace phenomena in the Venus dimer, extending the time-dependent simulation of the excitonic interaction to higher energy levels would offer a more comprehensive view of non-equilibrium dynamics in fluorescent proteins.

{Several aspects of the environmental treatment are deliberately simplified and bound the quantitative accuracy of $J$. First, the dielectric model treats the surroundings through bulk-water parameters ($\varepsilon_{\rm eq}\approx 78$, $\varepsilon_{\rm opt}\approx 1.78$, $\tau_D\approx 8.3$\,ps); a chromophore enclosed by the $\beta$-barrel is only partially solvent-exposed, so the relevant static permittivity is lower and spatially heterogeneous, and the tight inter-barrel packing makes the intervening solvent distribution asymmetric rather than the isotropic continuum assumed in the point-dipole coupling [Eq.~(\ref{eq:J_PD})] and the effective dielectric-relaxation model. These therefore represent approximations whose effect should be probed with a position-dependent or atomistic, explicitly polarisable embedding~\cite{ozaydin2023unraveling, mennucci2012}, which can introduce sizeable additional screening at this separation. We emphasise that the environment's fast electronic polarisability is itself already represented, at the continuum level, by the optical dielectric $\varepsilon_{\rm opt}$ that screens the coupling; what such an embedding would add is the explicit, position-dependent atomistic response of the protein and solvent, rather than electronic polarisability per se. Crucially, however, the central result (the coupling imprinted at the instant of absorption, $J(0)$) is set by the optical dielectric $\varepsilon_{\rm opt}$ alone: sweeping the static permittivity $\varepsilon_{\rm eq}$ over $4$--$78$ (protein interior to bulk water) leaves $J(0)$ exactly invariant and rescales only the long-time, fully relaxed limit. The protein-versus-water dielectric question therefore bears on the late-time damping of the coupling, not on the magnitude of the Davydov splitting set at absorption. Second, the single-exponential Debye form omits the fast librational and intramolecular components of the water response; these act on tens of femtoseconds and would provide a partial polarisation prior to full nuclear reorganisation, so the optical limit is best regarded as an upper bound on the early-time coupling. We note more broadly that the breakdown of the point-dipole approximation reported here is the expected behaviour whenever the chromophore separation is comparable to the molecular dimensions~\cite{sobakinskaya2018theory, scholes2003}; the present case quantifies its magnitude for a suspended fluorescent-protein dimer.}

Further refinements to increase the accuracy of our simulation estimates are possible, though they would incur a significant computational cost. Such refinements include enlarging the QM region and treating the phenolic hydrogen fully quantum mechanically to capture proton tunnelling \cite{chakraborty2008development,bourne2024quantum}. Additionally, future work will extend this framework to incorporate explicit vibronic coupling and evaluate the impact of non-Markovian protein phonons on long-term coherence dynamics.

\section{Data and Code Availability}

All numerical results are reproduced by a single orchestrator, \texttt{reproduce\_paper.py}, available on GitHub: \url{https://github.com/rchristie95/TerachemVenusA206}. It drives the structural build and QM/MM setup (\texttt{qmmm\_\allowbreak tddft\_\allowbreak pipeline.py}), the ORCA TDDFT reference, the ORCA DLPNO-STEOM-CCSD site energy and transition density, the Q-Chem EOM-CCSD(fT)/ADC(2) validation, and the coupling analysis, writing all reported quantities to a summary file. The repository also provides the analysis tools: \texttt{coupling\_core.py} (transition-density I/O and the GPU Coulomb kernel), \texttt{run\_nvt.py} (the audited production trajectory), \texttt{coupling\_dcd\_steom.py} (full-grid GPU coupling for every saved frame), \texttt{align\_\allowbreak steom\_\allowbreak density.py} (placing the STEOM density in the dimer frame), \texttt{multipole\_analysis.py} (multipole breakdown of the coupling), \texttt{export\_\allowbreak dipole\_\allowbreak geometry.py} (the two placed transition dipoles and centroids for the CD couplet), \texttt{make\_tandem\_panels.py} (Fig.~\ref{fig:tandem}), and \texttt{open\_\allowbreak quantum\_\allowbreak dynamics.py} (open-quantum-system dynamics, which regenerates the dynamics figures and the $T_2^*$ and dielectric sensitivity sweeps). Electronic-structure calculations used ORCA 6.1 \cite{neese2020orca,neese2022orca} and Q-Chem \cite{epifanovsky2021qchem}; classical setup used OpenMM/PDBFixer \cite{openmm2017,pdbfixer}; densities were rendered in PyMOL \cite{delano2002pymol}.

\section{Acknowledgements}
R.C and J.J. acknowledge support from the Institute for Information \& Communications Technology Promotion (IITP) grant funded by the Korea government (MSIP) (No. 2019-000003). Y.K. is grateful for support from the G-LAMP Program of the National Research Foundation of Korea, funded by the Ministry of Education (RS-2024-00445180). We thank Ian T. Abrams and Esme-Somerset Gregory for stimulating discussions. We also acknowledge the TeraChem package \cite{terachem2009,seritan2021terachem}, whose GPU-accelerated TDDFT enabled rapid exploratory screening of the chromophore's excited states during the discovery phase of this work.

\bibliography{Bibliography}

\begin{algorithm*}[ht]
\caption{Quantum-Classical Hybrid Excitonic Coupling Pipeline \\(\texttt{reproduce\_paper.py})}
\label{alg:pipeline}
\footnotesize
\begin{algorithmic}[1]
\Procedure{PipelineMain}{}
    \State \Call{Stage1\_ClassicalSetup}{}
    \State \Call{Stage2\_TDDFT}{}      \Comment{ORCA: single-excitation reference}
    \State \Call{Stage3\_STEOM}{}      \Comment{ORCA: correlated site energy + density}
    \If{validation requested}          \Comment{Stage 4 is optional}
        \State \Call{Stage4\_Validate}{} \Comment{Q-Chem: EOM-CCSD(fT)/ADC(2)}
    \EndIf
    \State \Call{Stage5\_TDC}{}         \Comment{static and thermal-ensemble coupling}
\EndProcedure

\vspace{0.2cm}
\Procedure{Stage1\_ClassicalSetup}{}
    \State Initialise system with PDBFixer (add missing atoms/hydrogens)
    \State Load AMBER14 XML forcefield parameters
    \State Solvate both dimer and monomer systems
    \State Relax both solvated geometries via OpenMM classical minimisation, with equilibrium bulk dielectric constant ($\varepsilon_{\rm eq}$)
    \State Partition system into QM and MM regions (link-atom boundary)
    \State Evaluate formal charges via CCD CIF parsing
    \State Extract MM point charges, applying short-range electrostatic damping
    \State Export monomer QM region coordinates as .xyz file
\EndProcedure

\vspace{0.2cm}
\Procedure{Stage2\_TDDFT}{}
    \State Load the $44$-atom QM region and its MM point-charge field (shared with Stage 3)
    \State Configure ORCA TDDFT ($\omega$B97X-D3/6-311G**, Tamm--Dancoff, point-charge embedding)
    \State Execute TDDFT energy calculation
    \State Parse excited-state roots and select the bright state by highest oscillator strength
    \State Reconstruct the transition density from natural transition orbitals (method-consistency coupling check)
\EndProcedure

\vspace{0.2cm}
\Procedure{Stage3\_STEOM}{}
    \State Build the reduced (CR2 + Tyr203) QM region with MM point-charge field
    \State Run ORCA DLPNO-STEOM-CCSD (def2-SVPD, RIJCOSX, TightPNO)
    \State Verify convergence and deterministic active space; parse bright state
    \State Reconstruct $\rho^{g\to e}$ from natural transition orbitals; scale to $f$
\EndProcedure

\vspace{0.2cm}
\Procedure{Stage4\_Validate}{} \Comment{optional}
    \State Run Q-Chem EOM-CCSD, EOM-CCSD(fT) and ADC(2) on the gas-phase chromophore
    \State Confirm STEOM matches the triples-corrected benchmark
\EndProcedure

\vspace{0.2cm}
\Procedure{Stage5\_TDC}{}
    \State Load the relaxed dimer frame and the STEOM (or TDDFT) transition density
    \State Rigid-fit (Kabsch) the density onto each chromophore site $L,R$
    \State \Call{CalculateCouplingGPU}{Site L, Site R} at the minimised geometry
    \State Repeat over the tandem-dimer NVT ensemble for the distribution
\EndProcedure

\vspace{0.2cm}
\Procedure{CalculateCouplingGPU}{$\mathbf{r}_L, q_L, \mathbf{r}_R, q_R$}
    \State Initialise Numba CUDA or PyOpenCL context
    \State Initialise $J = 0$
    \For{each grid point $i$ in Site A accumulate:}
        \State Convert reciprocal distance ($1~\text{\AA}^{-1} = 0.529177~a_0^{-1}$)
        \State $J \gets J + \sum_j \frac{q_{L,i} \cdot q_{R,j}}{|\mathbf{r}_{L,i} - \mathbf{r}_{R,j}|}$
    \EndFor
    \State Apply optical dielectric screening factor ($\varepsilon_{\mathrm{opt}}$)
    \State \Return Excitonic Coupling $J$
\EndProcedure
\end{algorithmic}
\end{algorithm*}

\end{document}